\documentclass[floats,floatfix,nofootinbib]{revtex4}
\usepackage[dvips]{graphicx}
\usepackage{graphics}
\usepackage{amssymb}
\usepackage{bm}
\usepackage{amsmath}
\usepackage{color}
\usepackage{epstopdf}

\begin{document}

\title{Stellar Instability from Parametric Resonance}

\author{Rodrigo Maier\footnote{rodrigo.maier@uerj.br},
\vspace{0.5cm}}

\affiliation{Departamento de F\'isica Te\'orica, Instituto de F\'isica, Universidade do Estado do Rio de Janeiro,\\
Rua S\~ao Francisco Xavier 524, Maracan\~a,\\
CEP20550-900, Rio de Janeiro, Brazil\\
}


\date{\today}

\begin{abstract}
In this paper we examine the stability of stellar configurations in which the interior solution is described by a closed
FLRW geometry sourced with a charged pressureless fluid
and radiation. An interacting vacuum component and a conformally coupled massive scalar field are also included. Given a simple factor for the energy transfer between the pressureless fluid and the vacuum component we obtain bounded interior oscillatory solutions. We show that in proper domains of the parameter space the interior dynamics is highly unstable so that the break of the KAM tori leads to a disruptive ejection of mass. For such configurations the interior solution asymptotically matches an exterior Reissner-Nordstr\"om-de Sitter spacetime.
\end{abstract}

\maketitle

\section{Introduction}
\label{intro}

The issue of an interacting dark energy in deep IR as in high UV has been a subject of interest over the last years.
In fact, from a cosmological point of view it has been shown that apart from relieving
some cosmological tensions of observational data\cite{sal}-\cite{kumar}, 
an interacting vacuum component may give rise to nonsingular models\cite{marco}.
In the framework of black hole physics, it has been shown that
Yukawa black holes\cite{maiery} or nonsingular Reissner-Nordstr\"om-de Sitter spacetimes\cite{maierns}
may be obtained if one considers proper interacting vacuum components.
In this context, a question which naturally arises 
is what would be the consequences of assuming an interacting dark energy in gravitational collapse processes which may generate stable/unstable stars. 

The problem of the gravitational collapse in General Relativity has been
object of several important works. In the realm of black hole formation, the seminal paper due to Oppenheimer and Snyder\cite{Oppenheimer:1939ue} furnished an
interior solution for a Schwarzschild spacetime assuming the
collapse of a spherically symmetric cloud of nonrelativistic and pressureless particles.
As an extension of this model, Vaidya made the inclusion of radiation in the exterior spacetime\cite{Vaidya:1951zz}.
On the other hand, excluding the presence of radiadion, Misner and Sharp made important progress
considering the gravitational collapse of a matter distribution more realistic than dust\cite{Misner:1964je}.
In a sequel of their work, a simplified heat-transfer process was introduced engendering an outward flux of neutrinos
\cite{Misner:1965zza}. Further analysis due to Chan et al. (see \cite{chan} and references therein) have studied the case of anisotropic gravitational collapse models. Recently, a proper examination of the stability of
neutron stars with a more realistic equation of state was performed in \cite{Pretel:2020mji}.
In this context, it is well known that the stellar structure in hydrostatic equilibrium
is governed by the Tolman-Oppenheimer-Volkoff (TOV) equations. For the case of a non-perfect fluid,
TOV equations were extended\cite{Sharma:2007hc} in order to include pressure anisotropy.
In this framework of stellar structure and evolution, two typical behaviours have deserved attention in the last decades.
It is understood that internal mechanical forces, thermal instabilities or turbulent motions
may drive oscillating internal waves which depend on the star interior properties\cite{samadi}.
The propagation of such waves produces an oscillating power spectrum
of the modes which may furnish important information about the stellar structure\cite{garcia}.
On the other hand, luminous stellar explosions regarded as supernovae (SNe) refer to the final stage of
massive stars in which the progenitor object collapses either to a neutron star, a black hole or 
is completely destroyed. Although the observational behaviour of these events is well understood\cite{Inserra:2019ciq,Jha:2019svc,Modjaz:2019flw},
a proper explanation about the mechanisms that trigger SNe ejection of mass remains uncertain.
In this paper we propose a simple inceptive model in which a conformally coupled massive scalar 
field may account to such a behaviour.

We organize the paper as follows. In Section $2$ we present the interior dynamics
of a Friedmann star in which the matter content is given by a charged pressureless fluid
and radiation. We show how bounded interior oscillatory solutions may be obtained once 
an interacting vacuum component and a conformally coupled massive scalar field are also assumed.
In Section $3$ we discuss the exterior spacetime which asymptotically corresponds to a
Reissner-Nordstr\"om-de Sitter geometry for proper configurations. Finally, in Section $4$ we leave our final remarks.

\section{Interior Dynamics}
\label{sec:1}

We start by considering the Einstein field equations
\begin{eqnarray}
\label{eq1}
G_{\mu\nu}=\kappa^2(T_{\mu\nu} - V_Ig_{\mu\nu})
\end{eqnarray}
where $G_{\mu\nu}$ is the Einstein tensor and $\kappa^2\equiv 8\pi G$.
The energy-momentum tensor $T_{\mu\nu}$ is constructed assuming that the matter content
of the model is given by a charged dust fluid, radiation and a 
conformally coupled massive scalar field. That is:
\begin{eqnarray}
T_{\mu\nu} = {^{(d)}}T_{\mu\nu}+{^{(\gamma)}}T_{\mu\nu}+{^{(\phi)}}T_{\mu\nu},
\end{eqnarray}
where ${^{(d)}}T_{\mu\nu}$ and ${^{(\gamma)}}T_{\mu\nu}$ stand for the energy-momentum tensors
of the charged dust fluid and radiation, respectively. The former can be written\cite{vickers} as
\begin{eqnarray}
\label{cdust}
{^{(d)}}T_{\mu\nu}=\rho_d u_\mu u_\nu + \sigma M_{\mu\nu},
\end{eqnarray}
where $\sigma$ is a negative coupling constant ($\sigma\propto -1/4\pi$) and
\begin{eqnarray}
M_{\mu\nu}=F_{\mu}^{~~\alpha}F_{\nu\alpha}-\frac{1}{4}g_{\mu\nu}F_{\alpha\beta}F^{\alpha\beta},
\end{eqnarray}
with $F_{\mu\nu}\equiv \nabla_\nu A_\mu-\nabla_\mu A_\nu$ as the Faraday tensor.
The radiation component, on the other hand, reads
\begin{eqnarray}
{^{(\gamma)}}T_{\mu\nu}=\frac{\rho_\gamma}{3}(4 u_\mu u_\nu+ g_{\mu\nu}).
\end{eqnarray}
Taking into account the lagrangian for a conformally coupled massive scalar field 
\begin{eqnarray}
\label{eq2}
{\cal L}_\phi=-\frac{1}{2}\Big[\phi_\alpha\phi_\beta g^{\alpha\beta}+m^2\phi^2+\frac{1}{6} R\phi^2\Big],
\end{eqnarray}
its respective energy-momentum tensor is given by
\begin{eqnarray}
{^{(\phi)}}T_{\mu\nu}= \phi_{,\mu}\phi_{,\nu}+{\cal L}_\phi g_{\mu\nu}+\frac{1}{6}\Big[\Box{(\phi^2)}g_{\mu\nu}+R_{\mu\nu}\phi^2-(\phi^2)_{,\mu,\nu}\Big].
\end{eqnarray}
Finally, we denote by $V_I$ a vacuum component
which interacts with the charged dust fluid. Such interaction is described by
an energy-momentum 4-vector $Q_\nu$ so that the
Bianchi identities furnish
\begin{eqnarray}
\label{eq5m1}
\nabla_\mu {^{(d)}}T^\mu_{~~\nu}=Q_\nu=\nabla_\nu V_I.
\end{eqnarray}

Let us now consider a FLRW interior geometry in comoving coordinates $(r, \theta,\varphi)$ given by
\begin{eqnarray}
\label{eq6}
ds^2=-dt^2+a^2(t)\Big[\frac{dr^2}{1-kr^2}+r^2(d\theta^2+\sin^2{\theta}d\varphi^2)\Big],
\end{eqnarray}
where $t$ is the time coordinate, $a(t)$ the scale factor and $k$ the $3$-curvature. 
From the conservation equations 
$\nabla_\mu{^{(\gamma)}}T^{\mu}_{~~\nu}=0$ 
we then obtain
\begin{eqnarray}
\rho_\gamma=\frac{E_\gamma}{a^4},
\end{eqnarray}
where $E_\gamma$ is a positive constant of integration.
On the other hand, the equation of motion for an homogeneous scalar field reads
\begin{eqnarray}
\label{eq111012}
\nabla_\mu{^{(\phi)}}T^{\mu}_{~~\nu}=0 \rightarrow\ddot{\phi}+3H\dot{\phi}+\Big[\frac{k}{a^2}+ 2H^2+\dot{H}+m^2\Big]\phi=0.
\end{eqnarray}
In the case of spherical symmetry, the only independent nonvanishing component of $F_{\mu\nu}$ is $F_{tr}=F(t,r)$.
Therefore, the Einstein field equations (\ref{eq1}) can be written as
\begin{eqnarray}
\label{ef1}
H^2+\frac{k}{a^2}=\frac{\kappa^2}{3}\Big[\rho_d+\rho_\gamma+V_I+\sigma(1-kr^2)\frac{ F^2(t,r)}{2a^2}+\frac{\dot{\phi}^2}{2}+H\dot{\phi}\phi+\frac{\phi^2}{2}\Big(H^2+\frac{k}{a^2}+m^2\Big)\Big],\\
\nonumber
\dot{H}+\frac{3H^2}{2}+\frac{k}{2a^2}=\frac{\kappa^2}{2}\Big\{V_I-\frac{\rho_\gamma}{3}+\sigma(1-kr^2)\frac{ F^2(t,r)}{2a^2}~~~~~~~~~~~~~~~~~~~~~~~~~~~~~~~~~~~~~~~~~~~~~~\\
\label{ef2}
~~~~~~~~~~~-\frac{\dot{\phi}^2}{6}+\frac{1}{2}(H^2+m^2)\phi^2+\frac{1}{3}\Big[\ddot{\phi}\phi+2H\phi\dot{\phi}+\Big(\dot{H}+\frac{k}{2a^2}\Big)\phi^2\Big]
\Big\}.
\end{eqnarray}
Imposing homogeneous energy densities together with an homogeneous vacuum component, we end up with
the condition
\begin{eqnarray}
F(t, r)=\frac{N(t)}{\sqrt{1-kr^2}}.
\end{eqnarray}
Substituting (\ref{cdust}) in (\ref{eq5m1}) we then obtain 
\begin{eqnarray}
\label{eq1612}
\nabla_\mu {^{(d)}}T^\mu_{~~\nu}=\nabla_\mu(\rho_d u^\mu u_\nu)+\sigma\nabla_\mu M^\mu_{~~\nu}=Q_\nu.
\end{eqnarray}
At this stage one may assume that $Q_\nu={\cal Q}_\nu+J_\alpha F_{\nu}^{~~\alpha}$ where $J^\alpha=\epsilon u^\alpha$ is a $4$-current with
$\epsilon$ being the density of electric charge.
Employing the Maxwell equations we then obtain
\begin{eqnarray}
\label{eqmax}
\sigma \nabla_\mu M^\mu_{~~\nu}=\sigma\nabla_\mu (F^{\mu}_{~~\alpha})F_{\nu}^{~~\alpha}=J_\alpha F_{\nu}^{~~\alpha},
\end{eqnarray}
Making $\nu=t$ in (\ref{eqmax}) we obtain
\begin{eqnarray}
N(t)=\frac{N_0}{a},
\end{eqnarray}
where $N_0$ is a constant. On the other hand, for $\nu=r$, equation (\ref{eqmax}) furnishes
\begin{eqnarray}
\epsilon(t, r)=2 \sigma N_0\Big(\frac{\sqrt{1-kr^2}}{ra^3}\Big).
\end{eqnarray}
As the physical radius $R$ of the matter distribution is proportional to the scale factor $a$ for a constant comoving 
radius $r$, from the above we note that $\epsilon$ scales as $R^{-3}$, as one should expect. 
Nevertheless, at a first glance one might identify a problem in the above charge density profile since it diverges as 
$r\rightarrow 0$. However, given the spherical symmetry of such matter distribution one should expect that
the overall charge should be spread out only in a small neighbourhood of the surface. In a more realistic model
this interior solution could be interpreted as a thin Friedmann layer in a small neighbourhood of the surface
to be matched with a metric which describes a more involved stellar core -- an issue to be addressed in a future work.

To proceed, in order to assure that the interior matter distribution bounces when a minimum $3$-volume
is reached,
we shall now assume that the energy-momentum $4$-vector ${\cal Q}_\nu$ has the following covariant prescription\cite{marco,maierns}
\begin{eqnarray}
\label{eq14}
{\cal Q}^\mu =\frac{4}{3}   (V_0-V_I) (\nabla_\alpha u^\alpha)u^\mu.
\end{eqnarray}
In the above, $V_0$ is a positive constant. 
Substituting (\ref{eq14}) in (\ref{eq1612}) we then obtain 
\begin{eqnarray}
\label{r01}
\nabla_\mu (\rho_d u^\mu u_\nu)=\nabla_\mu V_I=\frac{4}{3}   (V_I-V_0) (\nabla_\alpha u^\alpha)u^\mu.
\end{eqnarray}
A straightforward integration of the differential equations (\ref{r01}) furnishes
\begin{eqnarray}
V_I=V_0+\frac{\lambda}{a^4},\\
\rho_d=\frac{E_d}{a^3}-\frac{4\lambda}{a^4}.
\end{eqnarray}
where $\lambda$ and $E_d$ are positive constants of integration. Therefore, Einstein equations (\ref{ef1}) and (\ref{ef2}) read
\begin{eqnarray}
\label{fii}
H^2+\frac{k}{a^2}=\frac{\kappa^2}{3}\Big[\frac{E_d}{a^3}+\Big(\frac{2 E_\gamma+N_0^2\sigma-\lambda}{2a^4}\Big)+V_0+\frac{\dot{\phi}^2}{2}+H\dot{\phi}\phi+\frac{\phi^2}{2}\Big(H^2+\frac{k}{a^2}+m^2\Big)\Big],\\
\nonumber
\dot{H}+\frac{3H^2}{2}+\frac{k}{2a^2}=\frac{\kappa^2}{2}\Big\{V_0-\Big(\frac{2E_\gamma-3N_0^2\sigma-6\lambda}{6a^4}\Big)~~~~~~~~~~~~~~~~~~~~~~~~~~~~~~~~~~~~~~~~~~~~~\\
~~~~~~~~~~~-\frac{\dot{\phi}^2}{6}+\frac{1}{2}(H^2+m^2)\phi^2+\frac{1}{3}\Big[\ddot{\phi}\phi+2H\phi\dot{\phi}+\Big(\dot{H}+\frac{k}{2a^2}\Big)\phi^2\Big]\Big\}.
\end{eqnarray}

It is now useful to rewrite the first integral (\ref{fii}) and the equation of motion of the scalar field (\ref{eq111012}) in terms of the so-called conformal time $d\eta=a^{-1}dt$ together
with the rescaling $\psi\equiv \kappa a \phi$. In this case, the Friedmann equation (\ref{fii}) for a closed metric reads
\begin{eqnarray}
\label{eqn4}
3{a^\prime}^2+W(a)=\kappa^2 (E_\gamma-3 V_1) +\frac{1}{2}[{\psi^\prime}^2+(1+m^2 a^2)\psi^2],
\end{eqnarray}
where primes denote derivatives with respect to conformal time, $V_1\equiv \lambda-\sigma N_0^2/6$ and
\begin{eqnarray}
\label{eqn5}
W(a)=3a^2-\kappa^2\Big(V_0a^4+E_d a\Big).
\end{eqnarray}
The equation of motion of the scalar field $\psi$, on the other hand is given by
\begin{eqnarray}
\label{pieq}
\psi^{\prime\prime}+(1+m^2a^2)\psi=0.
\end{eqnarray}

We are
now in a position to define a dynamical system equivalent
to equations (\ref{eqn4})--(\ref{pieq}):
\begin{eqnarray}
\label{eqn7}
\psi^\prime &=&-p_\psi,\\
\label{eqn7pa}
a^\prime &=& p_a/6,\\
\label{eqn7nn}
p^\prime_{\psi}&=&(1+m^2a^2)\psi,\\
\label{eqn7n}
p^\prime_a&=&-\frac{dW}{da}+m^2a\psi^2.
\end{eqnarray}
In the above, $p_\psi$ and $p_a$ are the canonical momenta connected to the scalar field $\psi$ and the scale factor $a$, respectively. 
In fact, with the above definitions it is easy to show that (\ref{eqn4}) turns into a Hamiltonian first integral given by
\begin{eqnarray}
\label{ham}
{\cal H}=\frac{p^2_a}{12}+W(a)-\kappa^2 (E_\gamma-3 V_1) -\frac{1}{2}[p_\psi^2+(1+m^2 a^2)\psi^2]=0.
\end{eqnarray}
\begin{figure}
\begin{center}
\includegraphics[width=7.8cm,height=5cm]{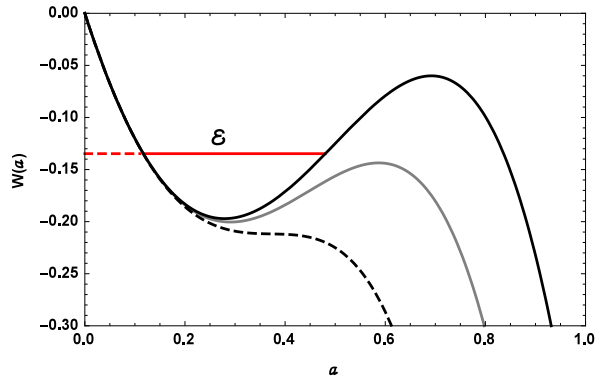}
\caption{The potential $W(a)$ for a closed model with $k=1$. In the above we have fixed $\kappa=1$ and $E_d=1.5$.
It is numerically shown that the potential $W(a)$ has two local extrema for $V_0=2.0$ (black curve) and $V_0=2.5$ (gray curve).
For $V_0=8k^3/(\kappa^3 E_d)^2$ (dashed curve) -- the upper limit for $V_0$ -- there are no extrema for the potential. 
Fixing $E_\gamma=0.01$, $V_1=0.052$, $p_{\psi 0}=0$ and $\psi_0=0.15$
we obtain ${\cal E}=-0.13475$ (red curve above).} 
\label{fig1}
\end{center}
\end{figure}

For $m =0$ the dynamical system (\ref{eqn7})--(\ref{eqn7n}) is separable, hence integrable.
In fact, from equations (\ref{eqn7}) and (\ref{eqn7nn}) we have a first integral
$E_{0\psi}= ({p_\psi^2} + \psi^2)/2$ which is a constant of motion.
It can be shown that 
the potential $W(a)$ has at most two local extrema (one local minimum $a_-$ and one local maximum $a_+$) for $a>0$
-- as long as $V_0<8k^3/(\kappa^3 E_d)^2$. Considering the surfaces
with energy ${\cal E}=\kappa^2 (E_\gamma-3V_1) +E_{0\psi}<0$ so that
$W(a_-)< {\cal E} < W(a_+)$ we see that the region $0<a<a_+$ is foliated by $2$-tori $S^1\times S^1$ which are the topological product of periodic
orbits of the separable sectors $(a, p_a)$ and $(\psi, p_\psi)$. 
Such $2$-tori trap the
dynamics in a finite region of the phase space and
${\cal E}$ is
a conserved quantity for those orbits. In the sector $(\psi, p_\psi)$ orbits have frequency $\nu_\psi=1/2\pi$
while in the sector $(a, p_a)$
\begin{eqnarray}
\label{eqn8}
\frac{1}{\nu_a}=2\int^{\beta_2}_{\beta_1}\sqrt{\frac{3}{{\cal E} -W(a)}}da.
\end{eqnarray}
Here, $\beta_1$ and $\beta_2$ are the two smaller real roots of ${\cal E} -W(a)$.
In Fig. 1 we show several plots of $W(a)$. 

A relevant question
which now arises is whether such tori ``survive'' once integrability is broken due to a nonvanishing mass $m$ for
the scalar field. In fact, assuming sufficiently small initial conditions
$(\psi_0,  p_{\psi 0})$, equation (\ref{pieq}) may be rewritten as
\begin{eqnarray}
\label{lame}
\psi^{\prime\prime}+(1+m^2a_0^2(\eta))\psi=0,
\end{eqnarray}
where $a_0(\eta)$ is the background solution for the scale factor
of the integrable dynamics with $m = 0$. Defining $\tilde{\nu}_{\psi}$ as the frequency in the
sector $(\psi, p_\psi)$ given by (\ref{lame}), a resonant behaviour 
will occur when the ratio
$R\equiv{\nu_{a}}/{\tilde{\nu}_\psi}$
is a rational number. Expanding $a_0(\eta)$ in (\ref{lame}), one can show that
\begin{eqnarray}
\label{feqp}
\tilde{\nu}_\psi\simeq\frac{1}{2\pi}\Big\{1+\frac{1}{2}\Big[m\frac{(\beta_1+\beta_2)}{2}\Big]^2-\frac{1}{8}\Big[m\frac{(\beta_1+\beta_2)}{2}\Big]^4\Big\}.
\end{eqnarray}
However, as the dynamics evolves the amplitude of the scalar field may grow so
that the solution of the integrable case $a_0(\eta)$ is no longer a good approximation to be
introduced in (\ref{lame}). This process may lead the dynamics
into a more unstable behavior, with the amplification of the resonance and
the break of the KAM tori\cite{Maier:2009zza,Maier:2013yh,arnold}. To analytically show this behavior, 
one may 
%
%
expand the non-integrable term of (\ref{ham}) in the action-angle variables $(\Theta_\psi=\tilde{\nu}_\psi\eta, {\cal J}_\psi, \Theta_a=\nu_a\eta, {\cal J}_a)$.
That is,
%
\begin{eqnarray}
\nonumber
\frac{1}{2}m^2a_0^2(\eta)\psi^2(\eta)=\frac{1}{2}m^2{\cal J}^{(0)}_a{\cal J}^{(0)}_\psi\sum_n[c_n\cos(2n\pi \Theta_a)]\cos(4\pi\Theta_\psi)
\end{eqnarray}
where $\psi(\eta)$ is an approximate solution of (\ref{lame}) and
$c_n$ are constant coefficients. The superior indexes in ${\cal J}_a$ and ${\cal J}_\psi$ denote that these
are the action variables for the integrable case. The Hamilton equation for ${\cal J}_a$ can then be integrated furnishing in its first approximation 
\begin{eqnarray}
\nonumber
{\cal J}_a\simeq \frac{1}{2}m^2{\cal J}^{(0)}_a{\cal J}^{(0)}_\psi\sum_n \frac{c_n}{2\pi n \tilde{\nu}_\psi}\Big[\frac{\cos(2\pi n\Theta_a-4\pi\Theta_\psi)}{{\nu_a}/{\tilde{\nu}_\psi}-2/n}
+\frac{\cos(2\pi n\Theta_a+4\pi\Theta_\psi)}{{\nu_a}/{\tilde{\nu}_\psi}+2/n}\Big].
\end{eqnarray}
From the above we see that the dominant resonance terms are those for which 
%
\begin{figure}
\begin{center}
\includegraphics[width=7.8cm,height=5cm]{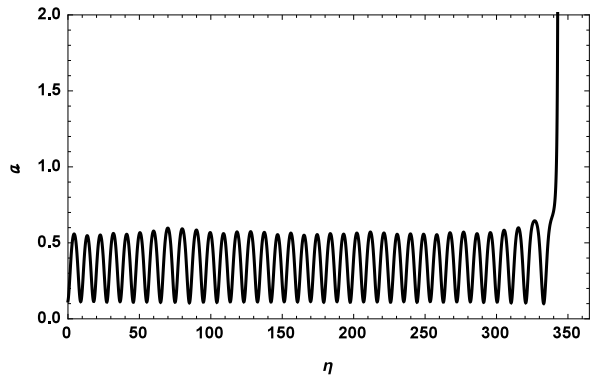}
\caption{The behaviour of the scale factor $a(\eta)$. For $\eta\simeq 343$ the scale factor diverges triggering a disruptive ejection of mass.}
\end{center}
\end{figure}

\begin{figure}
\begin{center}
\includegraphics[width=7.8cm,height=5cm]{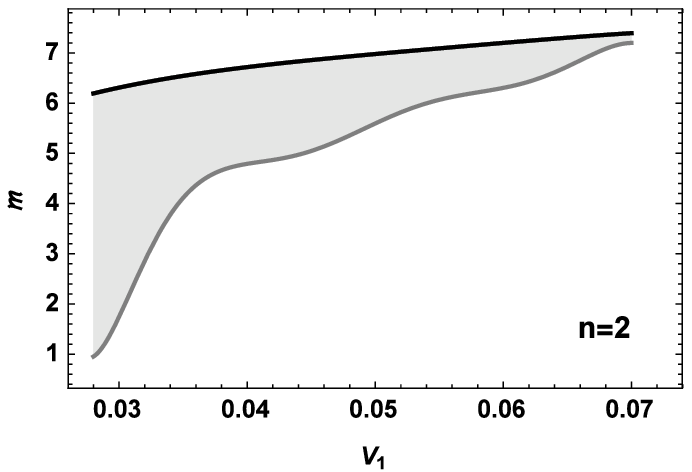}\includegraphics[width=7.8cm,height=5cm]{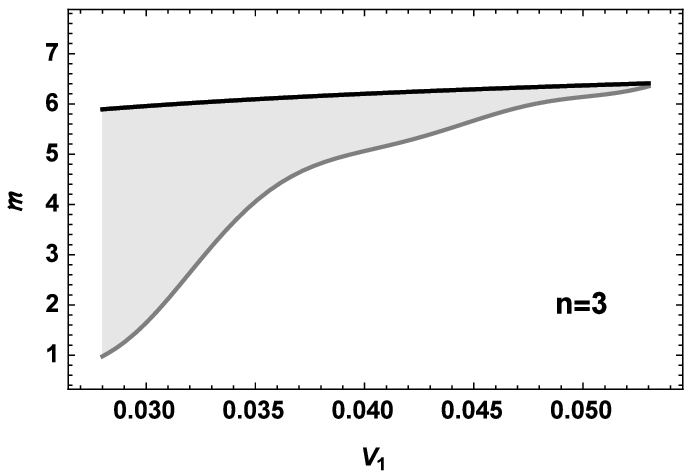}\\\includegraphics[width=7.8cm,height=5cm]{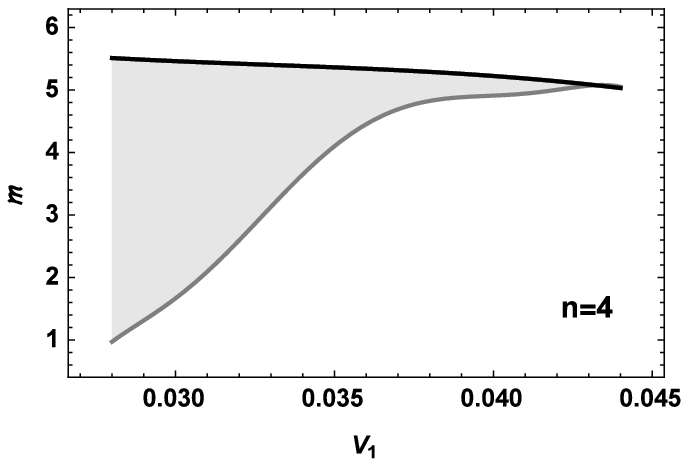}
\caption{Resonance domains for $n=2$ (top left panel), $n=3$ (top right panel) and
$n=4$ (bottom panel). Here we have fixed $\kappa=1$, $V_0=2.0$, $E_d=1.5$, $E_\gamma=0.01$ and initial conditions $p_{\psi 0}=0$, $\psi_0=0.15$ and $a_0=0.117$. In each dark solid line -- which were obtained from our analytical procedure 
due to (\ref{eqn8}), (\ref{feqp}) and (\ref{res}) -- the dynamics is highly unstable. There domain below such lines -- shaded areas -- is also resonant 
so that after a finite amount of time the scale factor diverges triggering a disruptive ejection of mass.
Below the gray lines we restore the domain of parametric stability analogous to that of the integrable case.} 
\label{fig1}
\end{center}
\end{figure}

\begin{eqnarray}
\label{res}
\frac{\nu_a}{\tilde{\nu}_\psi}\simeq\frac{2}{n}.
\end{eqnarray}
When such resonances occur one can
eventually obtain a loss of stability so that $a\rightarrow +\infty$ triggering a disruptive ejection of mass. 

To illustrate the above mentioned behaviour, let us consider the proper domain of the parameters
of Fig. 1 with $V_0=2.0$ (black curve). We also fix the initial conditions $p_{\psi 0}=0$, $\psi_0=0.15$ together with
$E_\gamma=0.01$, $V_1=0.052$ so that
${\cal E}=-0.13475$. The initial condition for the scale factor, $a_0\simeq 0.117$, is obtained from the first positive
root of $W(a)-{\cal E}$. 
For $n=3$, from (\ref{eqn8}), (\ref{feqp}) and (\ref{res}) we obtain $m\simeq 6.39$. 
Feeding the Hamiltonian constraint (\ref{ham}) with such parameters and initial condition we obtain the remaining initial condition 
$p_{a0}$.
Evolving the dynamical system imposing that the hamiltonian constraint is conserved, one can numerically show that
the scale factor diverges as $\eta\simeq 343$ triggering a disruptive ejection of mass. In Fig. 2 we illustrate this behaviour.

It is worth mentioning that there is a whole domain in the parametric space $(V_1, m)$ in which this unstable behaviour is manifest. In Fig. 3 we illustrate some examples of such domains for $n=2$ (top left panel), $n=3$ (top right panel) and
$n=4$ (bottom panel). Apart from $V_1$, $m$ and $p_{a0}$, we used the same parameters and initial conditions considered 
in Fig. 2. In Fig. 3 each dark solid line was obtained from our analytical procedure 
due to (\ref{eqn8}), (\ref{feqp}) and (\ref{res}) in order to find the respective resonances.
It can be numerically shown that the dynamics is highly unstable once one considers the parameters/initial conditions connected to these lines. There is also a whole domain below such lines -- shaded areas -- were the resonance mechanism is manifest
so that after a finite amount of time the scale factor diverges triggering a disruptive ejection of mass.
Below the gray lines we restore the domain of parametric stability analogous to that of the integrable case. It is also worth noting that one may obtain resonant configurations above the dark solid lines.
However, as our approximation 
is no longer valid for large masses -- above the dark solid lines --  one may safely regard the resonance domain as the shaded portions together with the black and gray lines.

\section{The Exterior Spacetime}
\label{sec:2}

We now consider the matching of the interior geometry with
the exterior spacetime. 
To this end let us assume that $(\bar{t}, \bar{r}, \bar{\theta}, \bar{\phi})$
are new coordinates defined by 
\begin{eqnarray}
\label{trans1}
\bar{t}:=\chi(\Gamma(t,\bar{r})),~~\bar{r}=ar,~~\bar{\theta}=\theta,~~\bar{\varphi}=\varphi.
\end{eqnarray}
The standard procedure to match the interior geometry with the exterior metric can be found in \cite{maierns,Oppenheimer:1939ue}.
Following the similar notation, we impose that the matching should be performed at the surface $r=\gamma={\rm constant}$,
so that
\begin{eqnarray}
\nonumber
g_{\bar{t}\bar{t}}=-\frac{1}{g_{\bar{r}\bar{r}}}\Big|_{r=\gamma}
\equiv
-\Big\{1-\frac{2GM}{\bar{r}}+\frac{\beta_q^2}{\bar{r}^2}-\frac{\kappa^2 }{3}\bar{r}^2\Big[V_0+L(\eta)\Big]\Big|_{r=\gamma}\Big\},
\end{eqnarray}
where
\begin{eqnarray}
L(\eta)=\frac{1}{2\kappa^2a^4}\Big[p_\psi^2+(1+m^2a^2)\psi^2\Big],
\end{eqnarray}
$M=4\pi\gamma^3 E_d/3$, $\beta_q^2=q^2 G/4\pi\epsilon_0$ and $q$ is the overall charge of the matter 
distribution\cite{maierns}.
At this stage, it is important to draw the reader's attention to a word to note. Let us consider the internal
oscillating bounded behaviour of the matter distribution -- with $\eta< 343$ as illustrated in Fig. 2, for instance. 
During this period the internal oscillating charged matter is responsible for an ejection of radiation making the exterior
spacetime stationary. In fact, such external radiation is needed in order to support the interior dynamics of the scalar field -- through the $L(\eta)$ function. However, once the scale factor diverges --
for $\eta\simeq 343$, for example -- there is a disruptive ejection of mass and $L(\eta)$ vanishes as $a\rightarrow +\infty$. 
Such behaviour is illustrated in Fig. 4. 
In this sense, the interior solution asymptotically matches an exterior geometry given by the 
Reissner-Nordstr\"om-de Sitter spacetime and the exterior metric reads
\begin{figure}
\begin{center}
\includegraphics[width=7.8cm,height=5cm]{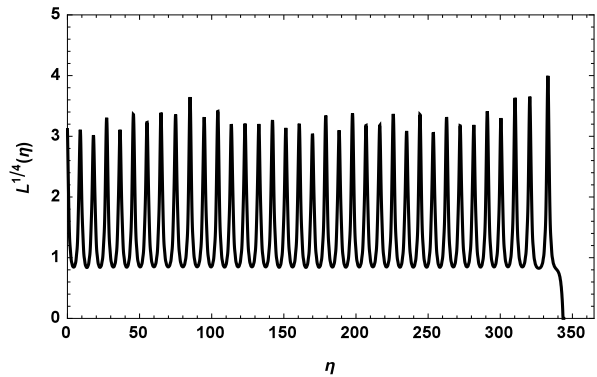}
\caption{The behaviour of $L(\eta)$ as a function of the conformal time. Here we see that $L(\eta)$ vanishes as $a\rightarrow +\infty$ at $\eta\simeq 343$.}
\end{center}
\end{figure}
\begin{eqnarray}
\label{extg}
ds^2=-F(\bar{r})d\bar{t}^2+\frac{1}{F(\bar{r})}d\bar{r}^2+\bar{r}^2(d\bar{\theta}^2+\sin^2{\bar{\theta}}d\phi^2),
\end{eqnarray}
where
\begin{eqnarray}
\label{grr}
F(\bar{r})\equiv \Big(1-\frac{2GM}{\bar{r}}+\frac{\beta_q^2}{\bar{r}^2}-\frac{\kappa^2 V_0}{3}\bar{r}^2\Big).
\end{eqnarray}
It is worth to note from (\ref{grr}) that $V_0$ plays the same role of a cosmological constant $\Lambda$. Therefore, assuming that $\kappa^2 V_0$ is sufficiently small it can be easily seen from (\ref{ham}) that 
$d^2a/dt^2 \simeq 0$ as $a\rightarrow +\infty$. Therefore, in the asymptotic regime the ejection of radiation completely ceases
making the exterior spacetime static as one should expect.

\section{Final Remarks}
\label{sec:2}

In this paper we propose a first analysis in which stellar stability may be connected to a conformally coupled massive scalar field. In order to assure that the matter distribution bounces when a minimum $3$-volume
is reached,
we assume that the internal pressureless matter interacts with vacuum component only through a covariant energy exchange\cite{marco,maierns}. In this case, bounded interior oscillatory solutions are obtained.
It is worth noting that the dynamics presented in this paper exhibit similar patterns as several bouncing cosmologies\cite{Maier:2009zza, Maier:2013yh}. In this sense, the interaction assumed in this paper plays just an effective role in order to make 
the dynamics nonsingular. Similar results -- i.e. the break of the KAM tori leading to a disruptive ejection of mass -- should be obtained for different bouncing models.

The obtention of the exterior stationary solution is a rather involved task which we intend to study in a further publication.
As mentioned above, such an exterior stationary metric is needed to support the interior scalar field dynamics.
For the case of stable configurations, the perpetual bounded oscillating interior spacetime could in principle be matched with
some sort of Vaidya spacetime\cite{Berezin:2017jsx}.
For the case of unstable configurations on the other hand, the same procedure could be performed
considering a Vaidya layer before its extension to the Reissner-Nordst\"om-de Sitter exterior solution\cite{vaidya}.

As a future perspective we also intend to consider the results of the present paper in order 
to furnish more realistic scenarios in which the oscillating scale factor may account for 
stellar internal waves together with a disruptive ejection of mass. The first step in this 
direction is a full examination of the resonance
domains furnishing constrains of more realistic parameters such as stellar masses. 
Another issue to be tackled is how the results shown in this paper fit
in
several models as neutrino heating, thermonuclear burning and magnetohydrodynamic instabilities which account to mass ejection in SNe (see \cite{Janka:2012wk} and references therein).

\end{document}